----------
X-Sun-Data-Type: default
X-Sun-Data-Description: default
X-Sun-Data-Name: YangMills1
X-Sun-Charset: us-ascii
X-Sun-Content-Lines: 620

\magnification=1200
\settabs 18 \columns

\baselineskip=12 pt

\def\sqr#1#2{{\vcenter{\vbox{\hrule height.#2pt
 \hbox{\vrule width.#2pt height#1pt \kern#1pt
 \vrule width.#2pt} \hrule height.#2pt}}}}

\def\operp{\hbox{${\kern+.25em{\bigcirc}
\kern-.85em\bot\kern+.85em\kern-.25em}$}}

\def\lsim{\;\raise0.3ex\hbox{$<$\kern-0.75em\raise-1.1ex\hbox{$\sim$}}\;}
\def\gsim{\;\raise0.3ex\hbox{$>$\kern-0.75em\raise-1.1ex\hbox{$\sim$}}\;}
\def\no{\noindent}

\def\ce{\centerline}
\def\ve{\vfill\eject}
\def\rdots{\mathinner{\mkern1mu\raise1pt\vbox{\kern7pt\hbox{.}}\mkern2mu
 \raise4pt\hbox{.}\mkern2mu\raise7pt\hbox{.}\mkern1mu}}

\def\e e{$e^+ e^-$ }



\rightline{UCLA/99/TEP/18}
\rightline{May 1999}
\vskip2.0cm

\ce{\bf GAUGED $q$-FIELDS}
\vskip.5cm

\ce{R. J. Finkelstein}
\vskip.3cm

\ce{\it Department of Physics and Astronomy}
\ce{\it University of California, Los Angeles, CA 90095-1547}
\vskip1.0cm

\ce{\it To Mosh\'e Flato}
\vskip1.0cm

\baselineskip=15pt
\line{{\bf Abstract.} \hfil}
\vskip.3cm

The straightforward description of $q$-deformed systems leads to transition
amplitudes that are not numerically valued.  To give physical meaning
to these expressions without introducing {\it ad hoc} remedies, one may
exploit an ``internal" Fock space already defined by the $q$-algebra.  This
internal space may be interpreted in terms of internal degrees of freedom 
of the deformed system or alternatively in terms of non-locality.  It is
shown that the $q$-deformation may give stringy characteristics to a
Yang-Mills theory.

\ve

\baselineskip=12pt
\line{{\bf 1. Introduction.} \hfil}
\vskip.3cm

Corresponding to the well known systems of quantum mehanics such as the
harmonic oscillator or the hydrogen atom there are $q$-systems obtained by
going over to quantum groups.  The $q$-systems have more degrees of freedom
than the system from which they are derived.  When the standard field
theories are similarly deformed the new degrees of freedom may be 
interpreted as expressions of non-locality by extended particles.  Here we
explore one aspect of this non-locality as it might appear in $q$-Yang-Mills
theories.
\vskip.5cm

\line{{\bf 2. The $q$-Yang-Mills Theory.} \hfil}
\vskip.3cm

Let $\psi(x)$ be the basis of the fundamental representation of a gauge
group.  Then if
$$
\psi^\prime(x) = T(x)\psi(x) \eqno(2.1)
$$
\no the covariant derivative $\nabla_\mu$ transforms as
$$
\nabla_\mu^\prime = T\nabla_\mu T^{-1} \eqno(2.2)
$$
\no so that
$$
\nabla_\mu^\prime\psi^\prime = T (\nabla_\mu\psi)~. \eqno(2.3)
$$

The corresponding gauge field $A_\mu$ is defined by
$$
A_\mu = \nabla_\mu-\partial_\mu~. \eqno(2.4)
$$
\no Then by (2.2)
$$
A_\mu^\prime = TA_\mu T^{-1} + T\partial_\mu T^{-1}~. \eqno(2.5)
$$
\no The field strength is
$$
F_{\mu\nu} = (\nabla_\mu,\nabla_\nu) \eqno(2.6)
$$
\no and transforms as
$$
F_{\mu\nu}^\prime = T F_{\mu\nu} T^{-1} \eqno(2.7)
$$
\no by (1.2).  The nature of the gauge group is so far unspecified.

We now assume that $T\epsilon SU_q(2)$.  Then
$$
T^t\epsilon T = T\epsilon T^t = \epsilon \eqno(2.8)
$$
\no where
$$
\epsilon = \left(\matrix{0 & q^{-1/2} \cr
-q^{1/2} & 0 \cr} \right)~. \eqno(2.9)
$$
\no Then, if $\psi$ is also a Dirac field,
$$
\psi^t C \epsilon \gamma^\mu\nabla_\mu\psi \eqno(2.10)
$$
\no is invariant where $\psi^t$ is the transpose of $\psi$ while $C$ and
$\epsilon$ satisfy
$$
\eqalignno{L^t C L &= C & (2.11) \cr
T^t \epsilon T &= \epsilon & (2.12) \cr}
$$
\no where $L$ is a Lorentz transformation and $C$ is the charge conjugation
matrix.  The basic Lagrangian may be chosen to be
$$
S = \int d^4x\biggl[-{1\over 4}{\rm Tr}_q F_{\mu\nu} F^{\mu\nu} +
i\psi^tC\epsilon\gamma^\mu\nabla_\mu\psi +{1\over 2}
[(\nabla_\mu\varphi)^t\epsilon
\nabla^\mu\varphi + \varphi^t\epsilon\varphi]\biggr] \eqno(2.13)
$$
\no where
$$
{\rm Tr}_qY = {\rm Tr}~QY \eqno(2.14)
$$
\no and
$$
Q = \left(\matrix{q^{-1} & 0 \cr 0 & q \cr} \right)~. \eqno(2.15)
$$
\no Here $\varphi$ is a Lorentz scalar that is also a fundamental representation
of $SU_q(2)$.

The first two terms of (2.13) are separately invariant and irreducible under
$SU_q(2)$.  The scalar contribution is invariant under only a subset of
$T$ transformations.
\vskip.5cm

\line{{\bf 3. The $SU_q(2)$ Algebra.} \hfil}
\vskip.3cm

Let
$$
T(0) = \left(\matrix{\alpha_0 & \beta_0 \cr
-q^{-1}\bar\beta_0 & \bar\alpha_0 \cr} \right)~. \eqno(3.1)
$$
\no Then (2.8) implies
$$
\eqalign{\alpha_0\beta_0 &= q\beta_0\alpha_0 \cr
\alpha_0\bar\beta_0 &= q\bar\beta_0\alpha_0 \cr
\beta_0\bar\beta_0 &= \bar\beta_0\beta_0 \cr} \qquad
\eqalign{&\alpha_0\bar\alpha_0 + \beta_0\bar\beta_0 = 1 \cr
&\bar\alpha_0\alpha_0 + q_1^2\bar\beta_0\beta_0 = 1 \cr
& q_1 = q^{-1}~. \cr} \eqno(3.2)
$$
\no Let us next introduce
$$
T(x) = \left(\matrix{e^{i\varphi(x)}\alpha_0 & e^{i\varphi(x)}\beta_0 \cr
-q_1e^{-i\varphi(x)}\bar\beta_0 & e^{-i\varphi(x)}\bar\alpha_0 \cr} \right) =
\left(\matrix{\alpha(x) & \beta(x) \cr
-q_1\bar\beta(x) & \bar\alpha(x) \cr}\right)~. \eqno(3.3)
$$
\no Then the elements of $T(0)$ and $T(x)$ satisfy the same relations
(3.2).  We shall write $(\alpha,\beta,\bar\beta,\bar\alpha)$ for the space
dependent matrix elements.

There is associated with the algebra (3.2) a state space.  Define the
ground state, $|0\rangle$, by
$$
\alpha|0\rangle = 0~. \eqno(3.4)
$$
\no Since $\beta$ and $\bar\beta$ commute we may require $|0\rangle$ to be
a common eigenstate of $\beta$ and $\bar\beta$.  Then
$$
\eqalignno{\beta|0\rangle &= b|0\rangle & (3.5) \cr
\bar\beta|0\rangle &= \bar b|0\rangle & (3.6) \cr}
$$
\no Note
$$
(\bar\alpha\alpha + q^2_1\beta\bar\beta)|0\rangle = |0\rangle \eqno(3.7)
$$
\no or
$$
|b|^2 = q^2~. \eqno(3.8)
$$

Define
$$
|n\rangle = \lambda_n \bar\alpha^n|0\rangle~. \eqno(3.9)
$$
\no Then
$$
\eqalign{\beta|n\rangle &= \lambda_n\beta\bar\alpha^n|0\rangle \cr
&= \lambda_n q^n\bar\alpha^n\beta|0\rangle \cr} \eqno(3.10)
$$
\no by iterating (3.2).  Then
$$
\beta|n\rangle = q^nb|n\rangle~. \eqno(3.11)
$$
\no Likewise
$$
\bar\beta|n\rangle = q^n\bar b|n\rangle~. \eqno(3.12)
$$
\no One also finds
$$
\lambda_n = \prod^{n-1}_0(1-|b|^2q^{2s})^{-1/2}  \eqno(3.13)
$$
\no Finally define the analogue of the Hamiltonian of the oscillator:
$$
H = {1\over 2}(\alpha\bar\alpha + \bar\alpha\alpha)~. \eqno(3.14)
$$
\no By (2.2)
$$
H = 1 - {1\over 2}(1+q^2_1)\bar\beta\beta~. \eqno(3.15)
$$
\no Then
$$
H|n\rangle = \biggl[1 - {1\over 2} (1+q^2)q^{2n}\biggr]|n\rangle \eqno(3.16)
$$
\no by (2.8).  Here the $|n\rangle$ are eigenstates of $\beta,\bar\beta$ and
$H$.  The levels depend on $q^{2n}$ rather than on $n$, i.e. they are
arranged in geometric rather than arithmetic progression and in this respect
resemble the $\langle n\rangle$ order of the $q$-oscillator.
\vskip.5cm

\line{{\bf 4. An Internal Space.}$^{(1)}$ \hfil}
\vskip.3cm

The existence of the algebra (3.2) leads to novel features of $q$-systems,
as one sees in the $q$-deformations of familiar elementary systems such as
the harmonic oscillator and the hydrogen atom.

In these examples one finds that the wave functions are not numerically
valued, but lie in the $q$-algebra.  To interpret these results according
to the usual rules, however, it is necessary to calculate numerically
valued transition probabilities between states represented by these
wave functions.  There is a natural procedure for doing this by
utilizing a Fock space associated with the algebra (3.2).  If such a path is
followed, however, these simple one-particle problems like the oscillator
and the hydrogen atom are endowed with internal degrees of freedom or partially
promoted to the complexity of quantum field problems.  In a similar way a
quantum field-theoretic problem will acquire a second Fock space or internal
degrees of freedom that can be given a non-local interpretation.  We shall
now see how this might work out for any $q$-deformed field theory, including
Yang-Mills.

A general field lying in the $q$-algebra will have the following expansion
$$
\psi_\mu(x) = \sum_\rho\bigl[f_\mu(\rho,x)a(\rho) + g_\mu(\rho,x)
\bar b(\rho)\bigr] \eqno(4.1)
$$
\no where
$$
\eqalignno{&\rho = (\vec p,r,s) & (4.2) \cr
&\sum_\rho = \biggl({1\over 2\pi}\biggr)^{3/2} \int
{d\vec p\over (2p_o)^{1/2}} \sum_{r,s} & (4.3) \cr
&f_\mu(\rho,x) = \sum_s U_\mu(\vec p,r,s)e^{-ipx}\tau_s & (4.4) \cr
&g_\mu(\rho,x) = \sum_s V_\mu(\vec p,r,s) e^{ipx}\tau_s & (4.5) \cr}
$$

Here $\mu$ is a generic tensor index, $\bar a$ and $\bar b$ are creation
operators for particles and antiparticles, the $r$-sum is the sum over
different polarizations and the $s$-sum is the sum over generators of the
$q$-algebra.

We may set
$$
\psi_\mu = \sum^4_{s=1} \psi_{\mu s}\tau_s \eqno(4.6)
$$
\no where the $\tau_s$ may be chosen as follows:
$$
\tau_s \equiv \tau_{ss^\prime} = T_{jk}\delta_{s,j}\delta_{s^\prime,k}~. 
\eqno(4.7)
$$
\no This choice of the $\tau_s$ implies a privileged gauge.  Any other
linear combination is allowed.  A general gauge transormation will carry
$\psi$ into a space where the basis elements are no longer linear in
$\tau_s$.
\vskip.5cm

\line{{\bf 5. Energy and Mass of a $q$-Scalar Particle.} \hfil}
\vskip.3cm

Let us consider the contribution of the scalar field in (2.13).  Then the energy operator is
$$
{\cal{E}} = \int T_{00} d\vec x = {1\over 2} \int
[(\nabla\varphi)^t\epsilon\nabla\varphi + 
m_0^2\varphi^t\epsilon\psi]d\vec x
\eqno(5.1)
$$
\no where
$$
\varphi = \biggl({1\over 2\pi}\biggr)^{3/2} \int{d\vec p\over (2p_0)^{1/2}}
\sum_s U(p,s)\bigl[e^{-ips}a(p,s) + e^{ipx}\bar a(p,s)\bigr]\tau_s
\eqno(5.2)
$$
\no with the usual commutators for the $a$ and $\bar a$.
Here the $s$-sum is over the generators of the $q$-gauge group.  We shall
compute just the contribution of the free field, i.e. without the vector
interaction.

Denote the basis states in Fock space by $|Nn\rangle$.  Then
$$
\bar a(\vec p,s) a(\vec p,s^\prime)\beta|N(\vec p,s)n\rangle =
N(\vec p,s)q^nb|N(\vec p,s)n\rangle \delta(s,s^\prime) \eqno(5.3)
$$
\no where $N(\vec p,s)$ is the population number of the state $(\vec p,s)$.

Now substitute (5.2) in (5.1) and use (5.3).  Then the eigenvalue of
${\cal{E}}$ is given by
$$
{\cal{E}}|Nn\rangle = \int d\vec p\sum_s U(\vec p.s)^t
\epsilon U(\vec p,s) p_o N(\vec p,s)\tau_s^2|Nn\rangle \eqno(5.4)
$$
\no where
$$
p_o^2 = \vec p~^2 + m^2~. \eqno(5.5)
$$
\no Choose
$$
\tau_s = \delta_{s1}\tau_1 \eqno(5.6)
$$
\no where
$$
\tau_1 = \beta~. \eqno(5.6b)
$$
\no By (5.4)
$$
{\cal{E}}|N(\vec p,1)n\rangle = \int d\vec p
\bigl[U(\vec p,1)^t\epsilon U(\vec p,1)\bigr] p_oN(\vec p,1)
\beta^2|N(\vec p,1)n\rangle~. \eqno(5.8)
$$
\no The factor $[U(p,1)^t\epsilon U(p,1)]$ 
is numerical and may be normalized to unity.
Then the eigenvalues of the partial energy contributed by the free
scalar field are
$$
{\cal{E}}^\prime = \int d\vec p N(\vec p,1) p_oq^{2n+1}~. \eqno(5.9)
$$
\no If $q$ is near unity, say $1 + {\epsilon\over 2}$, then the mass is
$$
\eqalign{m_oq^{2n+1} &=
 m_o\biggl(1 + {\epsilon\over 2}\biggr)^{2n+1} \cr
&= m_o\biggl(1+{1\over 2}\epsilon + n\epsilon\biggr) \cr} \eqno(5.10)
$$
\no for low $n$.
The mass lies between $m_o$ and infinity and becomes unbounded as $n$
becomes large.  It follows that a point particle with mass $m_o$ in a
$q=1$ theory becomes a particle with a mass spectrum and internal degrees
of freedom in a $q\not= 1$ theory.  For small values of $n$ the mass spectrum
resembles a string spectrum with the tension determined by $q$
and $m_o$.  If $q<1$ spectrum will be inverted and bounded.

One may construct similar arguments for the mass terms of other $q$-fields.
In every case, however, the actual dependence of the field on the $q$-algebra
is gauge dependent.  To make sense of this procedure one requires the
introduction of a privileged gauge, but that is also the way in which the
Higgs mechanism and the symmetry breaking vacuum completes the original
Yang-Mills theory.

Essentially equivalent to introducing an internal state space is averaging
over the $q$-group space with the aid of the Woronowicz integral.  In this
alternative formulation one replaces the action (2.13) which lies in the
$q$-algebra, and is therefore not numerically valued, by the following
average over the algebra:
$$
S = h \int d^4x~L \eqno(5.11)
$$
\no where $h$ stands for the Haar measure, or the Woronowicz integral, which
is a linear functional that may be evaluated term by term according to
$$
h[\alpha^s\beta^n\bar\beta^m] = \delta^{so}\delta^{mn}
{q^n\over [m+1]_q}~. \eqno(5.12)
$$
\no Thus $h$ projects out a power of $\beta\bar\beta$ that is evaluated as
the same power of $q$.  This result can be easily compared with the very
similar result obtained by the alternative procedure that employs the
state space.

The apparent similarity of a local gauge theory based on $SU_q(2)$ to one
based on $SU(2)$ is misleading, however, since only the latter has closure,
i.e. if $T_1$ and $T_2\epsilon SU(2)$ then $T_1T_2\epsilon SU(2)$; but if
$T_1$ and $T_2\epsilon SU_q(2)$ then $T_1T_2\epsilon \!\!\!/ SU_q(2)$ unless the
elements of $T_1$ commute with the elements of $T_2$.  In other words $T_1$
and $T_2$ must belong to different copies of the algebra.  For example, it
is possible that $T_1 = T(x_1)$ and $T_2 = T(x_2)$ where $x_1$ and $x_2$
both lie in spacetime but are not causally related; another possibility is
offered by a Kaluza-Klein theory if $x_1$ and $x_2$ do not both lie in
4-dimensional spacetime.  Such a collection of transformations may be
regarded as a non-local group.  

Alternatively one can assume that there are only two ``gauges" and that $T$
is a duality transformation.  One example of this kind of duality is realized
by the $q$-harmonic oscillator which has equivalent descriptions in terms of
either $(x,p)$ or $(a,\bar a)$.  These two representations are related by
the $q$-canonical transformation:
$$
\left(\matrix{a \cr \bar a \cr} \right) = T\left(\matrix{{i\over\hbar} p \cr
x \cr} \right) \qquad T\in SU_q(2)~. \eqno(5.13)
$$
\no Both sets satisfy $q$-commutation rules, namely:
$$
\eqalign{&a\bar a-q\bar aa = 1 \cr
&qxp - px = i\hbar \cr} \eqno(5.14)
$$
\no One may promote the oscillator to the role of a scalar field by 
introducing the conjugate fields $\pi(x)$ and $\psi(x)$ satisfying
$$
q\psi(x^\prime)\pi(x) - \pi(x)\psi(x^\prime) =
i\hbar\delta(x-x^\prime)~. \eqno(5.15)
$$

If the Fourier components of $\pi(x)$ and $\psi(x)$ are $p_k$ and $q_k$
and if we set
$$
\left(\matrix{a_k \cr \bar a_k \cr}\right) =
T \left(\matrix{{i\over\hbar} p_k \cr q_k \cr} \right) \eqno(5.16)
$$
\no then the field oscillators will satisfy $q$-commutators:
$$
a_k\bar a_{k^\prime} -q\bar a_{k^\prime}a_k = \delta_{kk^\prime}~.
\eqno(5.17)
$$
\no If one now calculates the energy of the free field one finds the
eigenvalues
$$
\langle n_k\rangle \hbar\omega_k \eqno(5.18)
$$
\no instead of
$$
n_k\hbar\omega_k~. \eqno(5.19)
$$
\no If $q$ is close to unity and we set $q=1+\epsilon$, then
$$
\eqalignno{\langle n\rangle &= {q^n-1\over q-1} & (5.20) \cr
{\langle n\rangle\over n} &= 1 + {n-1\over 2}~\epsilon & (5.21) \cr}
$$
\no Hence
$$
\langle n\rangle \hbar\omega = 
n\biggl[\hbar\omega + {n-1\over 2} \epsilon \hbar\omega\biggr]~. \eqno(5.22)
$$
\no This result may be interpreted to describe either a field with populations
of normal modes increasing as $\langle n\rangle$ with fixed mass or as
populations increasing linearly with $n$ but with increasing mass.
This example differs from the preceding case in that here there is no ``internal" quantum number and in addition the field commutators were normal
in the other case.  Both examples illustrate how the introduction of the
$q$-algebra may alter the particle spectrum.

To complete the discussion of this case it is again necessary to introduce
the internal space in order to compute transition probabilities.  Then the
field aquires an internal mass spectrum as in the preceding example.  In
both cases the internal $(\alpha,\bar\alpha)$ space is associated with the
algebra (3.2).  In the present case the $(\alpha,\bar\alpha)$ space results
from (5.17).  In the previous ($q$-Yang-Mills) case the dynamical fields lie
in the (3.2) algebra because of the postulated $SU_q(2)$ invariance.
\vskip1.0cm

\line{{\bf 6. Transition Amplitudes.} \hfil}
\vskip.3cm

We may consider the emission of a $q$-vector by a $q$-spinor induced by the
following interaction appearing in (2.13):
$$
\int i\psi^t(x)C\epsilon\gamma^\mu A_\mu(x)\psi(x) d\vec x \eqno(6.1)
$$
\no where
$$
A_\mu(x) = {1\over (2\pi)^{3/2}} \int {d\vec p\over (2p_o)^{1/2}}
\sum_p e_\mu(\rho)\bigl[e^{-ipx}A(\rho) + e^{ipx}\bar a(\rho)\bigr] \eqno(6.2)
$$
\no and
$$
e_\mu(\rho) = \sum_s e_\mu(\rho,s) \tau_s \qquad \rho = (\vec p,r)~. 
\eqno(6.3)
$$
\no Here $(\vec p,r)$ labels momentum and polarizatiion of the vector
particle while $\tau_s$ is to be summed over a selected portion of the internal
algebra generated by $(\alpha,\bar\alpha,\beta,\bar\beta)$.  Similarly we
set
$$
\psi(x) = \sum_s \psi_s(x)\hat\tau_s~. \eqno(6.4)
$$
\no Then the general matrix element of interest is
$$
\int d\vec x\langle N^\prime_\psi 
N_A^\prime n^\prime_\psi n^\prime_A|i\psi^t(x)C\epsilon\gamma^\mu A_\mu(x)
\psi(x)|N_\psi N_A n_\psi n_A\rangle \eqno(6.5)
$$
\no where $N$ and $n$ refer to the external and internal Fock spaces.

The emission of a specific vector particle $(\rho)$ by the fermionic source
will increase the corresponding population of the field from $N$ to $N+1$.
If a single fermion makes a transition from $n_\psi$ to $n^\prime_\psi$ in
the same event then the amplitude for the joint transition is
$$
\int d\vec x\langle n^\prime_\psi n_A^\prime|i\psi^t(x)
C\epsilon\gamma^\mu\psi(x)A_\mu(x)|n_\psi n_A\rangle (N_A+1)^{1/2} \eqno(6.6)
$$
\no or
$$
\int d\vec x\langle n_\psi^\prime|i\psi^t(x)C\epsilon\gamma^\mu\psi(x)|n_\psi\rangle
\langle n^\prime_A|A_\mu(x)|n_A\rangle(N_A+1)^{1/2}~. \eqno(6.7)
$$
\no In the $q=1$ theories there is no need of internal quantum numbers like
$n_\psi$ and $n_A$ since $\psi(x)$ and $A_\mu(x)$ ae numerically valued
in those theories.

Note that if $\tau_s$ or $\hat\tau_s$ is a function of $\beta$ and
$\bar\beta$ and does not contain free factors of $\alpha$ and $\bar\alpha$,
then there is no change in internal quantum number in the transition.  On
the other hand one has
$$
\langle n^\prime|\bar\alpha|n\rangle = \delta(n^\prime,n+1)
(1-q^{2n+2})^{1/2}~. \eqno(6.8)
$$

The ratio of the emission to the absorption probabilities of a vector
particle is
$$
R = {|\langle n+1|\bar\alpha|n\rangle|^2\over
|\langle n-1|\alpha|n\rangle|^2} =
{1-q^{2(n+1)}\over 1-q^{2n}} \eqno(6.9)
$$
\no multiplied by the corresponding factor for the spinor transition.

\vskip1.0cm

\line{{\bf References.} \hfil}
\vskip.3cm

\item{(1)} Finkelstein, R. Preprint UCLA/99/TEP/20.
\item{(2)} Woronowicz, S. L., Comm. Math. Phys. {\bf 112}, 125 (1989).

\bye